\documentclass[preprint,aps]{revtex4}
\usepackage{epsfig,latexsym,amssymb,amsmath,amsbsy,graphics,graphicx}

\def\bra{\langle}
\def\ket{\rangle}
\def\S{\boldsymbol{S}}

\begin{document}

\title{Dynamics of weakly coupled random antiferromagnetic quantum
spin chains}
\author{Eddy Yusuf and Kun Yang}
\affiliation{National High Magnetic Field Laboratory and
Department of Physics\\ Florida State University, Tallahassee, FL
32306}
\begin{abstract}
We study the low-energy collective excitations and dynamical
response functions of weakly coupled random antiferromagnetic
spin-1/2 chains. The interchain coupling leads to Neel order at
low temperatures. We use the real-space renormalization group
technique to tackle the intrachain couplings and treat the
interchain couplings within the Random Phase Approximation (RPA).
We show that the system supports collective spin wave excitations,
and calculate the spin wave velocity and spectra weight within
RPA. Comparisons will be made with inelastic neutron scattering
experiments on quasi-one-dimensional disordered spin systems such
as doped CuGeO$_3$.

\end{abstract}
\date{\today}
\maketitle



Antiferromagnetic (AF) quantum spin chains have been of interest
to physicists since the early days of quantum
mechanics\cite{bethe}. The one-dimensional nature of such systems
allowed for tremendous theoretical progress both in clean systems
by using exact solution and field theory
mapping\cite{bethe,haldane} and disordered systems within
renormalization group
framework\cite{mdh,fisher,hybg,hy,refael,damle}.
While such one-dimensional models have remarkably rich physics, in
general they do not give a complete description of real systems.
Real spin chain compounds, such as CuGeO$_3$\cite{hase} and
KCuF$_3$\cite{tennant}, always have some weak interchain couplings
present, which can change the physics at lowest
energy/temperature. For example, strictly one-dimensional models
do not exhibit phase transitions into states with broken symmetry,
while real 
spin chain systems often develop Neel order at very low
temperatures due to the weak (3D) interchain couplings. It is thus
important to study the effects of these interchain couplings to
fully understand the low-energy/temperature physics of real spin
chain compounds.

In this paper we study the low-energy collective excitations and
dynamical response functions of weakly coupled, disordered AF
spin-1/2 chains. Our work is motivated in part by the experimental
studies on doped CuGeO$_3$. In the absence of doping, it is a
spin-Peierls system in which the spins dimerize and form a gapped,
non-magnetic ground state. Upon doping, the system becomes
disordered, and both dimerization and spin gap get suppressed.
Amazingly, when doping reaches certain level the spins become Neel
ordered at low temperature, which has been observed experimentally
in Zn- and Si-doped CuGeO$_3$
\cite{hase2,oseroff,renard,regnault,kojima,hiroi,martin,katano,lussier,
shirane}. Since these experimental discoveries a number of
theoretical papers have addressed the static Neel ordering in
these
systems\cite{fukuyama1,melin1,melin2,mostovoy,dobry,melin3,melin4,melin5,
fukuyama2,yasuda1,affleck,yasuda2,joshi_kun} using mean-field
theory. On the other hand the collective excitations and dynamical
response functions, which have been studied experimentally using
inelastic neutron scattering\cite{shirane}, have not been studied
theoretically thus far.
The collective excitations and dynamical response functions are
the subjects of the present work. We go beyond mean-field theory
by allowing the Neel order parameter to fluctuate, and treat the
interchain coupling using the random phase approximation (RPA),
while tackle the intrachain coupling using the real space
renormalization group (RSRG) method\cite{mdh,fisher}. The RSRG
technique has been proven to be powerful in obtaining magnetic and
thermodynamics properties of random spin chains. Various numerical
techniques\cite{laflorencie,wessel,haas} have also been deployed
to study random chains and their results agree with those obtained
by RSRG.

Our strategy here is similar to that of Schulz\cite{schulz}, who
studied weakly coupled pure chains.
We find that despite the presence of disorder, the Neel state
supports linearly dispersing spin waves, in agreement with experiments. We
also obtain the spin wave velocity and the spectra weight of spin waves in
the dynamical response function in terms of microscopic
parameters of the system;
this allows for detailed quantitative comparison between theory and
inelastic neutron scattering experiments in the future.

Consider weakly coupled spin-1/2 antiferromagnetic chains with $z$
nearest neighbor for each chain. The suitable Hamiltonian to
describe this system is given by
\begin{equation}
\label{h0}
H=\sum_{i,\vec{n}}J_{i,\vec{n}}\S_{{i,\vec{n}}}\cdot\S_{{i+1,\vec{n}}} +
J_{\perp}\sum_{i,\vec{n},\vec{\delta}}\S_{{i,\vec{n}}}\cdot
\S_{{i,\vec{n}+\vec{\delta}}},
\end{equation}
where $i$ is the site index along the chain, $\vec{n}$ is the
chain index, and $\vec{\delta}$ is the index summed over the
nearest neighbors. The intrachain couplings $J_{i,\vec{n}}$ are
drawn from a random distribution function $P(J_{i,\vec{n}})$ (but
with independent realizations for each chain), while the
interchain coupling $J_{\perp}$ is taken to be constant. Both of
the intrachain and interchain couplings are taken to be positive.
Treating the interchain couplings in the mean-field
approximation\cite{schulz,joshi_kun}, the presence of these
couplings is effectively replaced by a staggered field which is
responsible for long-range ordering at low temperature. The mean
field approximation for the interchain couplings can be described
as follows : for a given site $i$, the staggered field acting on
this site is determined by the magnetization of neighboring sites
siting on neighboring chains. In general the staggered field
resulting from averaging the magnetization of the neighboring
sites will be random. However in the limit of infinite
coordination number $z$ (say in the limit of large
dimensionality), the number of neighboring sites contributing to
the staggered field at site $i$ becomes infinitely many and the
fluctuations are suppressed; the staggered field becomes uniform
in this limit because it becomes the average of the magnetization
over infinitely many neighboring sites (see below). With this
simplifying approximations at hand, the original problem is
reduced to a random AF spin chain in the presence of uniform
staggered field
\begin{equation}
\label{heff}
H_{1D}=\sum_{i}J_i\S_i\cdot\S_{i+1} - h\sum_{i}(-1)^iS_{i}^z.
\end{equation}
The staggered field $h$ is obtained through mean-field
self-consistency condition
\begin{equation}
\label{staggered}
h=zJ_{\perp}m,
\end{equation}
where $m$ is the disorder-averaged staggered magnetization. The
staggered magnetization at site $i$ is $m_i = (-1)^i \bra S_i
\ket$. The resulting effective 1D problem can be solved using the
real-space renormalization group (RSRG) method\cite{mdh,fisher},
from which the phase diagrams of the systems have been obtained
for various cases\cite{joshi_kun}.


In the present work we go beyond the static mean field approximation
and calculate the dynamical response functions
by treating interchain couplings within the Random
Phase Approximation (RPA), from which we also obtain the collective mode
spectrum of the system. Within RPA
the dynamic susceptibility is given by\cite{schulz}:
\begin{equation}
\label{rpa_sus}
\tilde{\chi}^{\alpha\beta}_{RPA} =
\frac{\tilde{\chi}^{\alpha\beta}_{1D}}{1-zJ_\perp\tilde{\chi}^
{\alpha\beta}_{1D}}
\end{equation}
where $\tilde{\chi}^{\alpha\beta}_{1D}$ is the disorder-averaged single chain
susceptibility matrix in the presence of the staggered field that satisfies
Eq. (\ref{staggered}).
The expression given above is valid for transverse and
longitudinal dynamic susceptibility. In this work we focus on the
transverse response function since the transverse part couples
more directly to the collective excitation of the system than the
longitudinal part does. Further motivation to study the transverse
susceptibility is provided by recent experiments which focus on
the transverse part; hence, $\alpha\beta = +-$ in Eq.
(\ref{rpa_sus}). It is worth noting that although we concentrate
on the transverse dynamic response in our calculation, the
formalism developed here can be readily applied to obtain the
longitudinal dynamic response and to study other random spin
chains.

Let us continue our discussion on the transverse dynamic response.
For a specific disorder configuration, the chain susceptibility
$\tilde{\chi}^{\alpha\beta}_{1D}$ represents the dynamical
response of the chain at wave vector $q'$ to an external
perturbation at wave vector $q$; in general $q'$ can be of any
value due to the presence of disorder which breaks the
translational symmetry. The symmetry is restored however once
disorder averaging is performed (in fact the system is
self-averaging), except for the doubling of unit cell by the
staggered field. Thus a perturbation with wave vector $q$ also
induces response at another wave vector $q'=q+\pi$, in addition to
the usual response at $q'=q$. Hence the chain transverse
susceptibility matrix is represented by a 2$\times$2 matrix :
\begin{equation}
\tilde{\chi}^{+-}_{1D}(q,\omega)=\left(
\begin{array}{ccc}
\chi^{+-}_{1D}(q,q,\omega) & & \chi^{+-}_{1D}(q,q+\pi,\omega)\\
\chi^{+-}_{1D}(q+\pi,q,\omega) & &\chi^{+-}_{1D}(q+\pi,q+\pi,\omega)\nonumber
\end{array}
\right).
\end{equation}
As a consequence of this, we can rewrite the RPA susceptibility, Eq.
(\ref{rpa_sus}) as follows :
\begin{equation}
\label{rpa_sus2}
\tilde{\chi}_{RPA}^{+-}(q,\omega)= \frac{1}{D (q, \omega)}
\left(
\begin{array}{ccc}
\chi_{11}(q, \omega) & & \chi_{12}(q, \omega)\\
\chi_{21}(q, \omega) & &\chi_{22}(q, \omega)
\end{array}
\right)
\end{equation}
where $D(q, \omega)$ is the determinant of
$1-zJ_\perp\tilde{\chi}^{+-}_{1D}(q,\omega)$.
Thus the calculation of $\tilde{\chi}_{RPA}^{+-}(q,\omega)$ reduces to the
calculation of $\chi_{1D}^{+-}(q, q', \omega)$.

To calculate $\chi_{1D}^{+-}(q, q', \omega)$, we use the RSRG method.
In the RSRG scheme, one assumes the two spins that are coupled by the
highest-energy bond form a ground state on this bond
(singlet when there is no field); this bond is eliminated
and
new effective bonds between the remaining spins are generated perturbatively;
this process is repeated
until the ultimate low-energy limit is reached. Within this scheme the single
chain response functions are approximated by the sum of those strongly
coupled bonds that form during
the RG process, and the coupling among different pairs of spins are neglected,
as outlined in Ref. \onlinecite{kedar}. This approximation is asymptotically
exact in the low-energy limit.
The difference between the
present case and that of Ref. \cite{kedar} is that in addition to the AF bonds
we also have the staggered field, which complicates the RG process. However it
has been shown earlier\cite{joshi_kun} that in the limit of weak field
(corresponding to weak interchain coupling), its effect on the RG flow is
negligible and we thus do not consider it here. We thus start by
considering a spin pair connected by a strong bond in the
presence of a staggered field whose Hamiltonian is given by :
\begin{equation}
\label{h2spin} H_0 = \Omega\S_1\cdot\S_2 + h (S_1^z-S_2^z),
\end{equation}
where $\Omega$ is the bond connecting the spin pair which we will
identify as the cutoff of the system at a given stage of RG and
$h$ is the uniform staggered field as defined in Eq. (\ref{staggered});
the eigen states of this Hamiltonian are :
\begin{eqnarray}
\label{spectrum}
|0\ket&=&1/\sqrt{1+C_0^2}(C_0|++\ket+|--\ket)\nonumber\\
|1\ket&=&1/\sqrt{1+C_1^2}(C_1|++\ket+|--\ket)\nonumber\\
|2\ket&=&|++\ket; ~|3\ket=|--\ket\,
\end{eqnarray}
with the corresponding energy eigenvalues :
\begin{eqnarray}
E_0&=&-\Omega/4 - \sqrt{(\Omega/2)^2+h^2}, \nonumber\\
E_1&=&-\Omega/4 + \sqrt{(\Omega/2)^2+h^2}, \nonumber\\
E_2&=& E_3=\Omega/4,
\end{eqnarray}
where the coefficients $C_0=2h/\Omega-\sqrt{1+(2h/\Omega)^2}$ and
$C_1=2h/\Omega+\sqrt{1+(2h/\Omega)^2}$. Equipped with the spectrum
of the spin pair, we proceed to calculate the dynamic structure
factor for the pair. The $T=0$ spectral representation of the
dynamic structure factor for a spin pair in the presence of
uniform staggered field is :
\begin{eqnarray}
&&S^{+-}_{pair}(q_1,q_2,\omega) =\sum_m
\bra0|S_{-q_1}^+|m\ket\bra m|S_{q_2}^-|0\ket\delta(\omega-\Delta E)\nonumber\\
&=& \frac{(C_0+e^{-iq_1l})(C_0+e^{iq_2l})}{1+C_0^2}
\delta(\omega-\Omega/2 - \sqrt{(\Omega/2)^2+h^2})\nonumber\\
\end{eqnarray}
where $S^\pm_q=S_1^{\pm}+e^{iql}S_2^{\pm}$ is the Fourier transform of the
spin operator, $l$ is the distance between two spins, $|0\ket$ is the ground
state and $|m\ket$ are the excited states $|++\ket$ and $|--\ket$ of
such pair as written explicitly in Eq. (\ref{spectrum}). The
system can only be excited to states different $m_z$ value because
the operator $S^\pm_q$ connects states with different $m_z$ value;
the transition can only happen from the ground state to the states
$|++\ket$ and $|--\ket$.

To calculate the dynamics structure factor of the whole chain, we
use the joint distribution function of bond-length and strength,
characterized in detail in Ref. \onlinecite{fisher}, and sum up
contributions of all the strongly coupled bonds that are formed
through the RSRG process. We follow the procedure outlined in Ref.
\onlinecite{kedar} closely to obtain the dynamic structure factor
of a single chain by summing the contribution from strongly
coupled pairs; the dynamic structure factor for a single chain is
given by :
\begin{eqnarray}
\label{1d} S_{1D}^{+-}(q_1,q_2,\omega) &=&
n_{\Gamma_\Omega}\int{dld\zeta
P(\zeta,l;\Gamma_\Omega)}S^{+-}_{pair}(q_1,q_2,\omega)\nonumber\\
\end{eqnarray}
where $P(\zeta,l;\Gamma_\Omega)$ is the joint distribution of
bond-length and strength. We follow the definitions outlined in
Ref. \onlinecite{fisher} to denote $n_{\Gamma_\Omega}$ as the
fraction of spins left at energy scale $\Gamma_\Omega$, $\zeta =
\ln(\Omega/J)$ as the log energy scale, $\Gamma_\Omega =
\log(\Omega_0/\Omega)$ as the log-cutoff of the energy scale,
$\Omega_0$ as the non-universal energy cutoff of the
original Hamiltonian and $\Omega$ as the energy cutoff of the
renormalized problem. The transverse susceptibility for a single
chain is then obtained by integrating the dynamic structure
factor:
\begin{equation}
\chi^{+-}_{1D}(q_1,q_2,\omega)=\int{\frac{d\omega'}{\pi}
\frac{S^{+-}_{1D}(q_1,q_2,\omega')}{\omega'-\omega-i\epsilon}},
\end{equation}
where $q_1, q_2 = q$ or $q+\pi$ and
$S^{+-}_{1D}(q_1,q_2,\omega)$ is given by Eq. (\ref{1d}).

We would like to study the collective excitations of the
system, whose spectrum is given by the singularity of
$\tilde{\chi}^{+-}_{RPA}$,
Eq. (\ref{rpa_sus}), or the condition $D(q, \omega)=0$.
The rotational invariance of the system combined with the self-consistent
condition Eq. (\ref{staggered}) requires
$D(q=0, \omega=0)=0$, because the staggered field can be rotated without
affecting the self-consistency.
As a consequence the energy of the
collective mode vanishes as the wave vector $q$ goes to zero. Thus
to obtain the collective mode dispersion it is natural to expand
the quantities $\chi^{+-}_{1D}(q,q,\omega),
\chi^{+-}_{1D}(q,q+\pi,\omega), \chi^{+-}_{1D}(q+\pi,q,\omega),$
and $\chi^{+-}_{1D}(q+\pi,q+\pi,\omega)$ for small $\omega$ and
$q$:
\begin{eqnarray}
\label{expansion} \chi^{+-}_{1D}(q,q,\omega) &\simeq&
\chi^{+-}_{1D}(0,0,0) + a_{q,q}\omega^2 +
b_{q,q}q^2\nonumber\\
\chi^{+-}_{1D}(q,q+\pi,\omega) &\simeq& a_{q,q+\pi}\omega +i
b_{q,q+\pi}q\nonumber\\
\chi^{+-}_{1D}(q+\pi,q,\omega) &\simeq& a_{q,q+\pi}\omega - i
b_{q,q+\pi}q\nonumber\\
\chi^{+-}_{1D}(q+\pi,q+\pi,\omega) &\simeq& \chi^{+-}_{1D}(\pi,\pi,0) +
a_{q+\pi,q+\pi} \omega^2\nonumber\\
&-& b_{q,q}q^2,
\end{eqnarray}
where the expansion coefficients are given by :
\begin{eqnarray}
\label{coef}
a_{q,q} &=& 2\int{\frac{d\omega'}{\pi}
\frac{S^{+-}_{1D}(0,0,\omega')}{\omega^{'3}}}\nonumber\\
b_{q,q} &=& -\frac{2l_v}{15}\int{\frac{d\omega'}{\pi}}
\frac{C_0}{1+C_0^2}\ln(\Omega_0/\Omega)\frac{1}{\omega'}
\frac{\omega^{'2}+h^{2}}{\omega^{'2}-h^{2}}\nonumber\\
a_{q,q+\pi} &=& 2\int{\frac{d\omega'}{\pi}
\frac{S^{+-}_{1D}(0,\pi,\omega')}{\omega'^2}}\nonumber\\
b_{q,q+\pi} &=& \frac{4}{3}\int{\frac{d\omega'}{\pi}}
\frac{C_0}{1+C_0^2}\frac{1}{\ln(\Omega_0/\Omega)}\frac{1}{\omega^{'2}}
\frac{\omega^{'2}+h^{2}}{\omega^{'2}-h^{2}}\nonumber\\
a_{q+\pi,q+\pi} &=& 2\int{\frac{d\omega'}{\pi}
\frac{S^{+-}_{1D}(\pi,\pi,\omega')}{\omega^{'3}}},
\end{eqnarray}
where
\begin{eqnarray}
S^{+-}_{1D}(0,0,\omega) &=& \frac{(1+C_0)^2}{1+C_0^2}\frac{1}{l_v
\ln^3(\Omega_0/\Omega)}\frac{1}{\omega}
\frac{\omega^{2}+h^{2}}{\omega^{2}-h^{2}}\nonumber\\
S^{+-}_{1D}(0,\pi,\omega) &=& \frac{1-C_0^2}{1+C_0^2}\frac{1}{l_v
\ln^3(\Omega_0/\Omega)}\frac{1}{\omega}
\frac{\omega^{2}+h^{2}}{\omega^{2}-h^{2}}\nonumber\\
S^{+-}_{1D}(\pi,\pi,\omega) &=&
\frac{(1-C_0)^2}{1+C_0^2}\frac{1}{l_v
\ln^3(\Omega_0/\Omega)}\frac{1}{\omega}
\frac{\omega^{2}+h^{2}}{\omega^{2}-h^{2}},\nonumber\\
\end{eqnarray}
where $l_v = a/(\overline{\ln(\Omega_0/J)})$ is the microscopic length scale
determined by the initial bond distribution; we use $\overline{x}$ to denote
the variance of x.
 Using the condition
that $D(q, \omega)= det[1-zJ_\perp\tilde{\chi}_{1D}^{+-}(q,\omega)]=0 $
for $\omega=q=0$, we
obtain quartic equation in $\omega$
\begin{equation}
A\omega^4 + B\omega^2 + C = 0,
\end{equation}
where the coefficients $A, B,$ and $C$ are :
\begin{eqnarray}
\label{coef2}
A &=& (zJ_\perp)^2 a_{q,q} a_{q+\pi,q+\pi}\nonumber\\
B &=& -zJ_\perp(1-zJ_\perp\chi^{+-}_{1D}(0,0,0))a_{q+\pi,q+\pi}\nonumber\\
&+&(zJ_\perp)^2((a_{q+\pi,q+\pi}-a_{q,q})b_{q,q}q^2 - a_{q,q+\pi}^2)\nonumber\\
C&=& (zJ_\perp(1-zJ_\perp\chi^{+-}_{1D}(0,0,0))b_{q,q}\nonumber\\
&-&(zJ_\perp)^2b_
{q,q+\pi}^2)q^2-(zJ_\perp)^2b_{q,q}^2q^4.
\end{eqnarray}
The solution to this quartic equation gives us the spin-wave
dispersion of the system. To the leading order of the wave vector
$q$ we obtain a linear dispersing spin-wave $\omega = v_s q$, with
\begin{equation}
\label{velocity}
v_s \simeq \sqrt{\frac{\pi}{90}} ~zJ_\perp m
\ln^{3/2}(\Omega_0/zJ_\perp m) l_v.
\end{equation}
In obtaining the result above for the spin-wave velocity, we have explicitly
worked in the limit $J_\perp \to 0$.

We now calculate the dynamic structure factor within RPA,
which can be accessed through Inelastic Neutron Scattering (INS)
experiment. To obtain the dynamic structure factor near $q=0$, we
take the imaginary part of the upper left component of the RPA
susceptibility matrix, Eq. (\ref{rpa_sus2}), i.e. $S_{RPA} =
\Im(\chi_{11}(q,\omega)/D(q,\omega))$,
\begin{equation}
S_{RPA}(q,\omega) = \Im\Big[\frac{\chi_{11}(q,\omega)}
{A\omega^4 + B\omega^2 + C -i\delta}\Big],
\end{equation}
where a small imaginary part, $i\delta$ has been introduced to
shift the pole in the determinant $D(q,\omega)$ to slightly above the
real axis.
The pole in the determinant
is realized when $\omega = v_s q$, where $v_s$ is the spin-wave
velocity given in Eq. (\ref{velocity}). A straightforward
calculation results in a simple form of the RPA dynamic structure
factor near $q=0$ :
\begin{equation}
\label{s_0}
S_{RPA} = \frac{1}{3} \frac{v_s q}{zJ_\perp} \delta(\omega-v_s q).
\end{equation}
Following the same procedure, we also obtain the dynamic structure
factor near $q=\pi+\delta q$, $S_{RPA} = \Im(\chi_{22}(q,\omega)/D(q,\omega))$.
 The result is as follows :
\begin{equation}
\label{s_pi} S_{RPA} = \frac{1}{2} \frac{zJ_\perp m^3
\ln^2(\Omega_0/zJ_\perp m)}{v_s (\delta q)} \delta(\omega-v_s
(\delta q)),
\end{equation}
where $m$ is disorder-averaged staggered magnetization and
$\Omega_0$ is non universal cutoff for the chain. Our calculation
predicts that a sharp peak develops at the pole where $\omega=v_s
q$. The intensity of the peak is proportional to the wave vector
$q (1/\delta q)$ near $q=0 (\pi)$. Within the Random Phase
Approximation framework we only get a sharp peak of the intensity,
as is shown by the delta function in the dynamic structure factor.
The peak shows more pronounced contribution from the dynamic
structure factor near $q=\pi$ because the long-range staggered
configuration is realized near this wave vector.
We comment here that while our expression for $v_s$ (Eq. (\ref{velocity})
involves parameters of the
random distribution ($\Omega_0$ and $l_v$)
that cannot be measured directly, our results on
the spectral weight can be compared directly with the intensity
of inelastic neutron scattering experiment, once $v_s$ is determined from the
measurement; this is because Eq. (\ref{s_0}) involve $v_s$ and other measurable
quantities only, and the same combination of $\Omega_0$ and and other
measurable quantities appears in Eq. (\ref{velocity}) and (\ref{s_pi})
($l_v$ is of order one lattice spacing for generic distributions). Thus our
results allow for detailed quantitative comparison with future experiments.


A few years earlier, Martin {\em et al.} studied the excitation
spectrum of doped CuGeO$_3$ using inelastic neutron
scattering\cite{martin}, and found sharp propagating spin wave
excitations when the system is Neel ordered, despite the fact that
the Neel phase is stabilized by disorder. They found the spin wave
spectrum to be linear. Our results agree with these experimental
findings, and it is clear that such propagating excitations must
be collective modes stabilized by interchain couplings, as single
random chains do not support such propagating modes\cite{kedar}.
In the present work we assume there is no dimerization, while in
doped CuGeO$_3$ dimerization survive and coexist with Neel order
when doping level is sufficiently low. It is straightforward to
generalize the present approach to the case with dimerization
\cite{hybg,joshi_kun}, as well as finite temperature and chains
with other spin sizes. Recently Masuda {\it et al.}\cite{masuda}
studied the dynamic spin-spin correlation of a new compound
BaCu$_2$(Si$_{1-x}$Ge$_{x}$)$_2$O$_7$ using inelastic neutron
scattering. This system can be described very well by
antiferromagnetic spin-1/2 chains with random exchange due to the
random distribution of Si and Ge atoms. The experimental data on
the dynamic structure factor on this compound fit the universal
scaling form predicted in Ref. \cite{kedar} very well. Our
theoretical work proposed here could be of relevance to this new
experimental realization of random exchange antiferromagnetic
spin-1/2 chain; in particular it would be interesting to study the
collective excitation on this compound in the ordered phase and
compare it with the results obtained here. We hope the present
work will motivate future experiments that will study the spectral
weights of the spin waves in detail, and test the predictions made
here on them.

We thank Kedar Damle for stimulating discussions. This
work was supported by NSF grant No. DMR-0225698, and the MARTECH.

\end{document}